\documentclass[conference,compsoc]{IEEEtran}
\author{
    \IEEEauthorblockN{
        Julia B. Kieserman\textsuperscript{1}, 
        Athanasios Andreou\textsuperscript{1}, 
        Laura Edelson\textsuperscript{2},
        Sandra Siby\textsuperscript{3}, 
        Damon McCoy\textsuperscript{1}, 
    }
    \IEEEauthorblockA{
        \textsuperscript{1}New York University \\
        \textsuperscript{2}Northeastern University \\
        \textsuperscript{3}New York University Abu Dhabi \\
        Emails: \{julia.kieserman, a.andreou, sandra.siby, mccoy\}@nyu.edu, l.edelson@northeastern.edu
    }
}

\usepackage{cite}
\usepackage{tabularx}
\usepackage{multirow}
\usepackage{comment}
\usepackage{graphicx}
\usepackage{xcolor}
\usepackage{url}
\usepackage{pifont}
\usepackage{booktabs}

\usepackage{caption}
\title{Reckless Designs and Broken Promises: Privacy Implications of Targeted Interactive Advertisements on Social Media Platforms}

\begin{document}

\maketitle
 
\begin{abstract}
    Popular social media platforms TikTok, Facebook and Instagram allow third-parties to run targeted advertising campaigns on sensitive attributes in-platform. These ads are interactive by default, meaning users can comment or ``react'' (e.g., ``like'', ``love'') to them. We find that this platform-level design choice creates a privacy loophole such that advertisers can view the profiles of those who interact with their ads, thus identifying individuals that fulfill certain targeting criteria. This behavior is in contradiction to the promises made by the platforms to hide user data from advertisers. We conclude by suggesting design modifications that could provide users with transparency about the consequences of ad interaction to protect against unintentional disclosure.
\end{abstract}

\section{Introduction}
Many social media companies own and operate advertising tools. With a wealth of users, and user data, they are well-equipped to help advertisers reach large numbers of people targeted across a diverse set of demographics and interests. There are known issues with targeted advertising, including that they exhibit discriminatory behavior, rely on sensitive user attributes, and target individual users~\cite{andreou2019measuring, speicher2018potential, gonzalez2021unique}. Nevertheless, social media companies claim that they protect user privacy from advertisers. For example, TikTok claims that they share only ``aggregated statistics and insights with users of our analytics services''\cite{tikTokPrivacyPolicy}. Similarly, Faceook promises not to tell ``advertisers who you are or sell your information to anyone''\cite{facebookPromise}. 

However, the design decisions made by platforms may impact their ability to uphold these privacy claims. TikTok and Meta, the focus of this study and two advertising platforms with significant user bases, have designed their infrastructure such that ads are interactive by default. Much like user-generated content, users can comment on or react to (e.g., ``like,'' ``love'') an ad. 
This leaves a record of engagement which, if made visible to advertisers, would create a privacy loophole; since advertisers already know the potentially sensitive nature of the targeting criteria they have selected (e.g., household income, dependents' age), this would allow them to learn the usernames of individuals that fit this criteria.

To understand if this is possible, we ran advertisements on TikTok and Meta (Instagram and Facebook). We found that neither platform takes precautions to obscure the individual usernames of those who interact with ads from advertisers. TikTok shares the username of individuals who comment on (but not react to) ad content directly through the advertising interface. Meta shares the usernames of both users who comment and users who react to an ad. This means that users who interact with targeted ads are inadvertently identifying themselves to advertisers as individuals who meet specific targeting criteria. In summary, we demonstrate that social media platforms leak the personal data of users who interact with advertisements directly to advertisers, in contradiction to their own documentation.

\begin{table*}[!t]
    \centering
    \renewcommand{\arraystretch}{1.5}
    \begin{tabular}{|c|c|c|c|}
    \hline
    \textbf{Platform} & \textbf{Person Who Commented Visible} & \textbf{Person Who Reacted Visible} & \textbf{Interactivity Configurable} \\
    \hline
    \textit{Meta} (Facebook + Instagram) & \ding{51} & \ding{51} & \ding{91} \\ \hline
    \textit{TikTok} & \ding{51} & & \ding{51} \\ \hline 
    \end{tabular}
    \caption{Summary interaction data visibility and configurability across platforms}
    \ding{91} The documentation suggests it is available in a limited capacity but it was not available to the papers' authors.
    \label{table:summary}
\end{table*}

\section{Background}
Many social media platforms provide an infrastructure and user interface for advertisers to run marketing campaigns.
These platforms make a set of design choices, and expose certain configuration options to advertisers, which dictate how users experience ads. As a baseline, advertisers typically need to specify a budget and creative copy (i.e., photo or video advertisements).  
After the initial setup, advertisers can use a provided user interface to view aggregated metrics about who viewed their advertisements, including demographics (e.g., age, gender, geography) and device information (e.g., iOS, Android).
Two of these design choices, when combined, have significant privacy implications: targeting and interactivity.

\paragraph{\textbf{Targeting}} Platforms like Meta and TikTok allow advertisers to choose from a set of platform-defined targeting criteria to determine what characteristics about users should make them eligible to view an ad. A subset of the targeting options are potentially sensitive including age and gender (TikTok and Meta), household income (TikTok and Meta), U.S. postal code (Meta), age range of dependents (Meta), and interest and behavior-based targeting (TikTok and Meta). Targeted advertising alone can be privacy invasive; prior work has shown it can facilitate discriminatory targeting~\cite{speicher2018potential}, use sensitive user attributes~\cite{andreou2019measuring}, and, with the right combination of attributes, target individual people~\cite{gonzalez2021unique}. Further, despite calls for user ad transparency and the introduction of features like Meta's ``Why Am I Seeing This Ad''~\cite{whyAd}, users still have no way to adequately understand why they were targeted for a particular ad~\cite{burgess2024seeing}, and thus what sensitive attribute groups they belong to. 

\paragraph{\textbf{Interactivity}} Ad interactivity, which allows users to comment on an ad, is a default behavior in both Meta and TikTok. It is configurable in TikTok, although the user interface hides the toggle and the accompanying text description discourages advertisers from disabling this feature (Appendix Figure~\ref{fig:tiktok-user-config}). Although the ability to disable comments in the Meta ads interface is reported to exist in a limited capacity, it was not available to our business account at the time of the study~\cite{metaDisableComments}. There are ways to limit interactions through the post, but not disable them entirely. 

\section{Case Studies: TikTok and Meta}
To find out if user information is visible to advertisers, we ran advertisements on TikTok and Meta using their UI business tools. Our advertisements were targeted to users over 18 in the United States.

\begin{figure}[ht]
    \centering
    \includegraphics[width=0.55\linewidth]{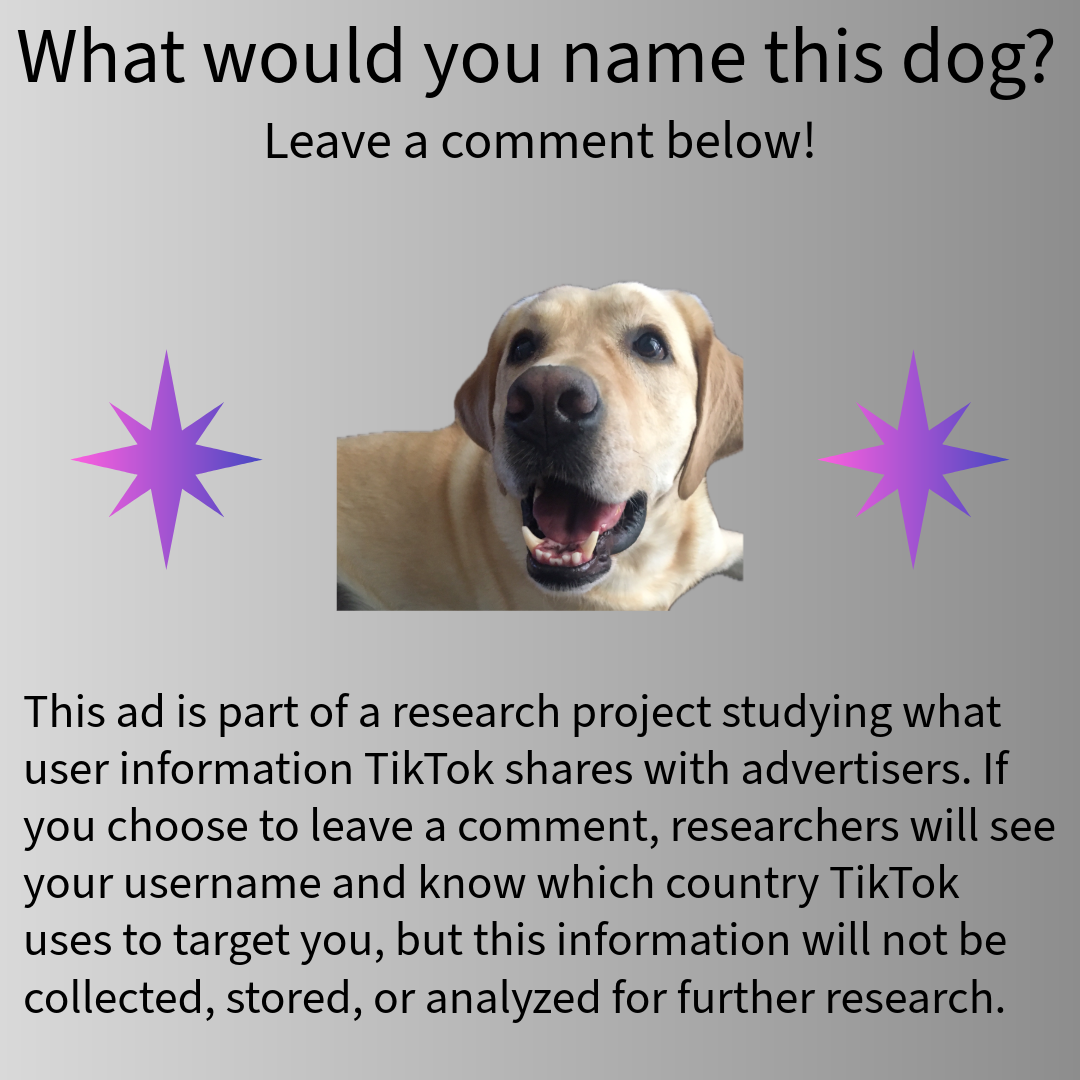}
    \caption{Advertisement run through TikTok's campaign platform, which targeted users over 18 in the United States.}
    \label{fig:tiktok-ad}
\end{figure}

\paragraph{\textbf{TikTok}}
We ran our TikTok advertisement (Figure~\ref{fig:tiktok-ad}) on December 18-20, 2025, where it was viewed 11,667 times.
Two users commented on the advertisement and one "liked" it. Through the TikTok advertising interface, we were able to view the name and profile image of the users that left a comment. We were not able to view the username of the individual who "liked" it. 

\paragraph{\textbf{Meta}}
We ran our Meta advertisement (nearly identical to Figure~\ref{fig:tiktok-ad}) on January 6-9, 2026, across Facebook and Instagram. It was shown to approximately 13,755 accounts on Facebook and 17,546 on Instagram. On Facebook, it received 4 reactions, all of which were visible to advertisers with a username and profile photo. On Instagram, it received 11 comments and 14 reactions. We were able to view the username and photo for 10 out of 11 comments and the username, account name, and profile photo for 8 of the 14 reactions. We do not know exactly why some of the comments/reactions were hidden, but we verified that the visible reactions and comments came from a mixture of public and private accounts. As our ad was not shared on either platform, we do not know how that action might impact data visibility but see it as an area for future work.

\paragraph{\textbf{Ethics}}
Our study was reviewed and approved by our Institutional Review Board (IRB). The content of the advertisement (Figure~\ref{fig:tiktok-ad}) clearly stated its intended purpose and the potential implications for users who interacted with it. We did not recruit participants and used the broadest possible targeting criteria through the platforms' interface to target only users over 18 who are in the United States. Finally, we did not remove any user data from the platform or perform further analysis.
\section{Discussion and Future Work}
Our findings show that users who interact with advertisements on TikTok and Meta expose their usernames to advertisers, thus revealing that they are identifiable by potentially sensitive targeting criteria. The platform decision to make ads interactive (by default or as the only option) is not simply an inevitable consequence of online advertising. Other platforms that serve advertisements to users (e.g., YouTube) treat it as fundamentally different content that users can not actively engage with and therefore are unlikely to have this privacy issue.

Platforms may argue that by choosing to interact with an ad, users are implicitly agreeing or even taking responsibility for sharing their username (or legal names~\cite{facebookNames}) with advertisers. However, they are doing more than that; they are linking their username to a specific set of unknown targeting criteria, likely unwittingly disclosing potentially sensitive information about themselves~\cite{andreou2018investigating}. Future work could investigate how users understand the privacy implications of interactive ads and how it changes their platform behavior.
 
Platforms like TikTok and Meta could consider other design choices that alleviate the risk of users unintentionally identifying themselves to advertisers. One method would be to include disclaimers in both the user documentation and in-platform ads, as well as a correction to the inaccurate statements currently in the documentation, that warn users of the consequences of interacting with an ad. However, this is not ideal as it puts undue burden on users. As an alternative, platforms could commit to turning interactivity off by default.

\section*{Acknowledgment}
This work was funded by National Science Foundation Grant \#2151837 and NYU Abu Dhabi's 19WSN Co-Mentoring Initiative.

\bibliographystyle{IEEEtran}
\bibliography{bibliography}

\clearpage
\appendix
\section{Appendix}

\begin{figure}[ht]
    \centering
    \includegraphics[width=0.85\linewidth]{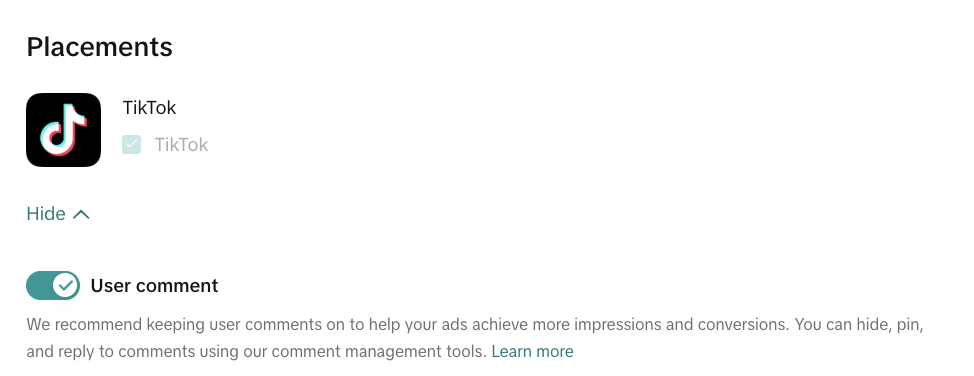}
    \caption{Configuration to disable user comments on TikTok advertisements}
    \label{fig:tiktok-user-config}
\end{figure}

\begin{figure}[ht]
    \centering
    \includegraphics[width=0.55\linewidth]{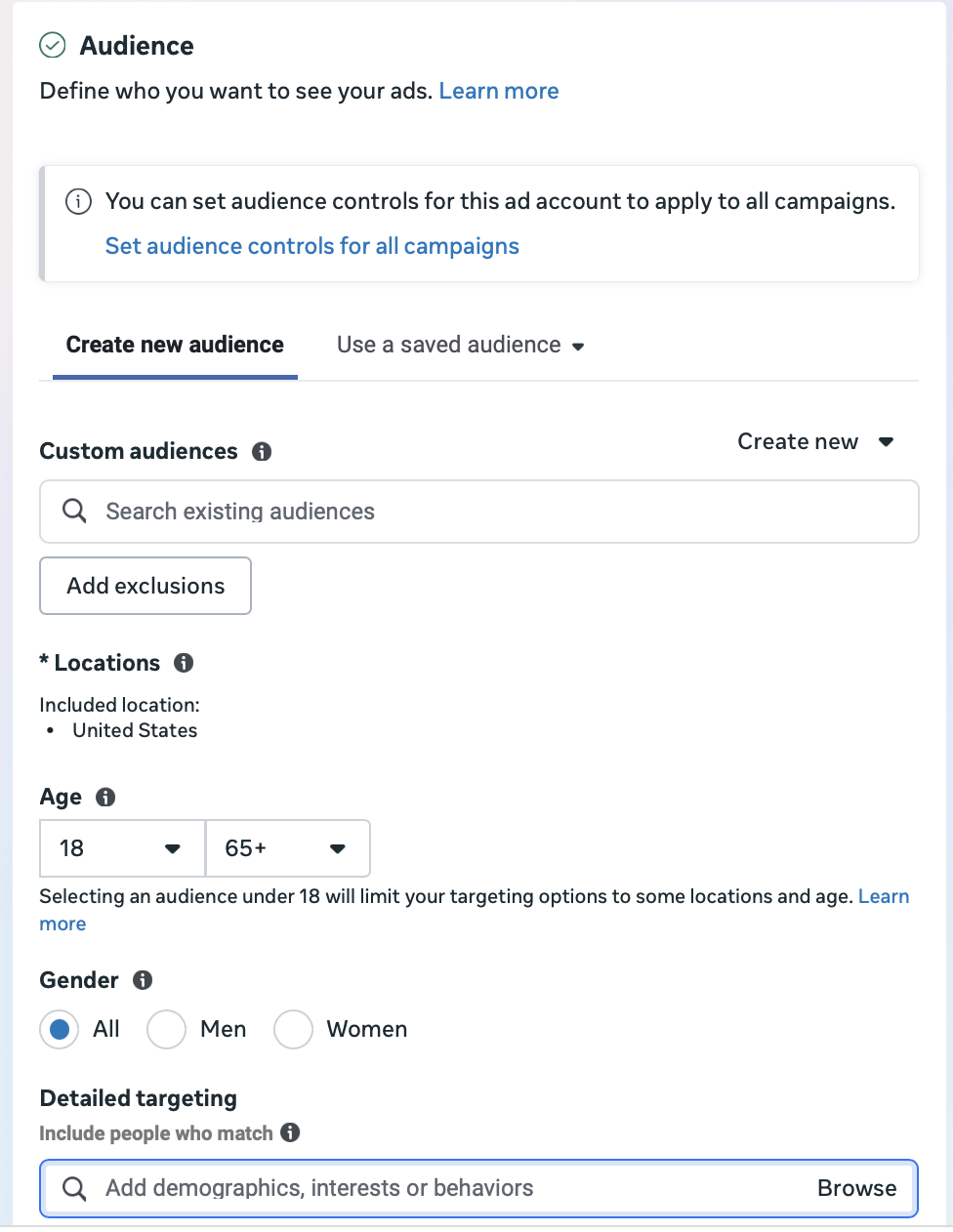}
    \caption{Screenshot of Meta's interface for setting targeting criteria}
    \label{fig:meta-targeting-UI}
\end{figure}

\end{document}